\documentclass[journal]{IEEEtran}

\usepackage[utf8]{inputenc}
\usepackage[T1]{fontenc}
\usepackage{hyperref}
\usepackage{url}
\usepackage{booktabs}
\usepackage{amsfonts}
\usepackage{amsmath}
\usepackage{amssymb}
\usepackage{mathtools}
\usepackage{graphicx}
\usepackage{xcolor}
\usepackage{microtype}
\usepackage{comment}
\usepackage{subcaption}      
\usepackage{multirow}
\usepackage{tabularx}
\usepackage{array}
\usepackage{colortbl}

\usepackage{algorithm}
\usepackage{algorithmic}

\usepackage[numbers, square, sort&compress]{natbib}

\newcommand{\agentforge}{\textsc{AgentForge}}
\newcommand{\swebench}{\textsc{SWE-bench}}

\title{AgentForge: Execution-Grounded Multi-Agent LLM Framework for Autonomous Software Engineering}

\author{
    Rajesh~Kumar, %
    \thanks{R. Kumar with International Research Center for Complexity Sciences, Hangzhou International Innovation Institute, Beihang University, Hangzhou, 311115, China(e-mail: \{rajakumarlohano@gmail.com}%
    Waqar~Ali, %
    \thanks{W. Ali is with the Department of Computer Science, College of Science, Mathematics and Technology, Wenzhou-Kean University, Wenzhou 325060, China (e-mail: waqar.uestc@yahoo.com).}%
    Junaid Ahmed,
     \thanks{Fakulti Teknologi Maklumat dan Komunikasi, 
Universiti Teknikal Malaysia Melaka, Melaka 76100, Malaysia (e-mail:  j.bhatti@iba-suk.edu.pk).}
    Najma Imtiaz Ali,
     \thanks{Computer Systems Engineering Department, Sukkur IBA University, Sindh , Pakistan (e-mail:  najma@utem.edu.my).}
Shaban Usman,
     \thanks{Yibin Park of University of Electronic Science and Technology of China, Yibin 644000, China (e-mail:  shabanusman@yahoo.com).}

}

\begin{document}

\maketitle

\begin{abstract}
Large language models generate plausible code but cannot verify correctness. Existing multi-agent systems simulate execution or leave verification optional. We introduce execution-grounded verification as a first-class principle: every code change must survive sandboxed execution before propagation. We instantiate this principle in \agentforge{}, a multi-agent framework where Planner, Coder, Tester, Debugger, and Critic agents coordinate through shared memory and a mandatory Docker sandbox. We formalize software engineering with LLMs as an iterative decision process over repository states, where execution feedback provides a stronger supervision signal than next-token likelihood. \agentforge{} achieves 40.0\% resolution on \swebench{} Lite, outperforming single-agent baselines by 26--28 points. Ablations confirm that execution feedback and role decomposition each independently drive performance. The framework is open-source at \url{https://github.com/raja21068/AutoCodeAI}.
\end{abstract}

\begin{IEEEkeywords}
Multi-agent systems, Large language models, Automated program repair, Execution verification, Software engineering, SWE-bench
\end{IEEEkeywords}

\section{Introduction}
\label{sec:intro}

Large language models (LLMs) perform well on code generation but remain unreliable for real-world software engineering. Practical tasks require reasoning over existing codebases, executing programs, and iteratively refining solutions under test feedback. Most current systems treat code generation as a single-step prediction problem, mapping a natural language description to a code completion~\cite{brown2020language}. This approach fails on tasks that require multi-file reasoning, test generation, and regression avoidance~\cite{chen2021codex}.

Recent work addresses these limitations with multi-agent systems that decompose development into roles such as planning, coding, testing, and reviewing. These systems improve performance on benchmarks such as SWE-bench through structured interaction between agents~\cite{trae2025}. Despite this progress, a key limitation remains: existing frameworks do not enforce grounded execution. They infer execution outcomes or rely on permissive environments, rather than observing actual program behavior.

This limitation is critical. Bug fixing requires a feedback loop plan, implement, execute, test, and revise driven by real execution signals. Without grounded feedback, models cannot reliably verify correctness or detect regressions. Simulated execution introduces systematic errors that propagate through the pipeline.

We introduce \textbf{AgentForge}, a multi-agent framework that enforces verified execution for autonomous software engineering. AgentForge decomposes bug fixing into five specialized agents: \textit{Planner}, \textit{Coder}, \textit{Tester}, \textit{Debugger}, and \textit{Critic}. The Planner generates a structured execution plan. The Coder produces minimal patches using unified diffs. The Tester synthesizes executable test cases. The Debugger iteratively repairs failures using execution feedback. The Critic validates the final result.

AgentForge grounds all decisions in two retrieval sources: (i) episodic memory of previously solved tasks and (ii) a live repository index of the current codebase. The system executes every generated patch inside a resource-constrained, network-isolated Docker sandbox. This design provides non-simulated execution feedback and enables a closed-loop Tester–Debugger cycle for iterative repair.
Table \label{tab:frameworks_axes} situates AgentForge among existing systems. Prior frameworks introduce role decomposition, knowledge graphs, or test-time scaling~\cite{khanzadeh2025agentmesh,tao2024magis,sgagent2026,trae2025}. Other approaches explore self-evolution and competitive reasoning~\cite{ecoevolve2026,swedebate2026}. None enforce mandatory sandboxed execution while combining dual retrieval with a full five-agent pipeline. AgentForge integrates these components into a unified, execution-grounded framework.

\begin{table*}[t]
\centering
\caption{Positioning Multi-Agent Frameworks Along Three Axes}
\label{tab:frameworks_axes}
\small
\setlength{\tabcolsep}{5pt} 
\renewcommand{\arraystretch}{1.25} 

\begin{tabular}{@{} l p{3.5cm} p{3.5cm} p{3.5cm} @{}}
\toprule
\textbf{Framework} 
  & \textbf{Execution Feedback} 
  & \textbf{Role Decomposition} 
  & \textbf{Memory/Retrieval} \\ 
\midrule

\textbf{AgentForge} 
  & Mandatory sandbox 
  & 5 roles (plan, code, test, debug, critic) 
  & Dual (episodic + repo) \\ 

\textbf{SWE-agent} 
  & Shell/ACI (optional) 
  & Single agent 
  & None \\ 

\textbf{OpenHands} 
  & Sandbox (optional) 
  & Single agent 
  & None \\ 

\textbf{Trae Agent} 
  & Test-time scaling 
  & 3 roles (gen, prune, select) 
  & None \\ 

\textbf{MAGIS} 
  & Implied 
  & 4 roles (mgr, custodian, dev, qa) 
  & None \\ 

\textbf{AgentMesh} 
  & None 
  & 3 roles (plan, debug, review) 
  & None \\ 

\textbf{Reflexion} 
  & Simulated 
  & Single + self-reflection 
  & Episodic memory \\ 

\bottomrule
\end{tabular}
\end{table*}

\textbf{Contributions.}
\begin{itemize}
    \item We formalize LLM-based software engineering as an \textit{execution-grounded iterative refinement problem}, where correctness is defined by external program execution rather than model-internal likelihood signals.

    \item We model this process as a sequential decision problem over repository states and cast it as an MDP, enabling analysis of feedback, credit assignment, and error propagation.

    \item We identify two key properties: (i) execution feedback provides a stronger supervision signal for functional correctness than next-token likelihood, and (ii) decomposing generation, testing, and debugging reduces error accumulation compared to monolithic self-repair.

    \item We instantiate these principles in \agentforge{}, a five-agent framework with structured orchestration, dual retrieval (episodic memory and repository index), and mandatory Docker-based execution.

\end{itemize}

\textbf{Novelty summary.} AgentForge is the first framework to mandate sandboxed verification for every code change, providing ground-truth execution feedback. It integrates five specialized agents (Planner, Coder, Tester, Debugger, Critic) with a dual-memory system (episodic memory + live repository index) – a combination absent from AgentMesh, MAGIS, SGAgent, and Trae Agent.

\textbf{Implications.} Our results indicate that verified execution feedback and structured pipeline design matter more than raw model scale for real-world software engineering. AgentForge provides an open-source baseline for future research in this direction.

The remainder of this paper is organized as follows. Section~\ref{sec:related} surveys related work. Section~\ref{sec:method} details the AgentForge architecture. Section~\ref{sec:experiments} describes the experimental setup and baselines. Section~\ref{sec:results} presents main results and ablations. Section~\ref{sec:conclusion} discusses limitations and future directions.

\begin{figure*}[ht]
    \centering
    \includegraphics[width=\linewidth]{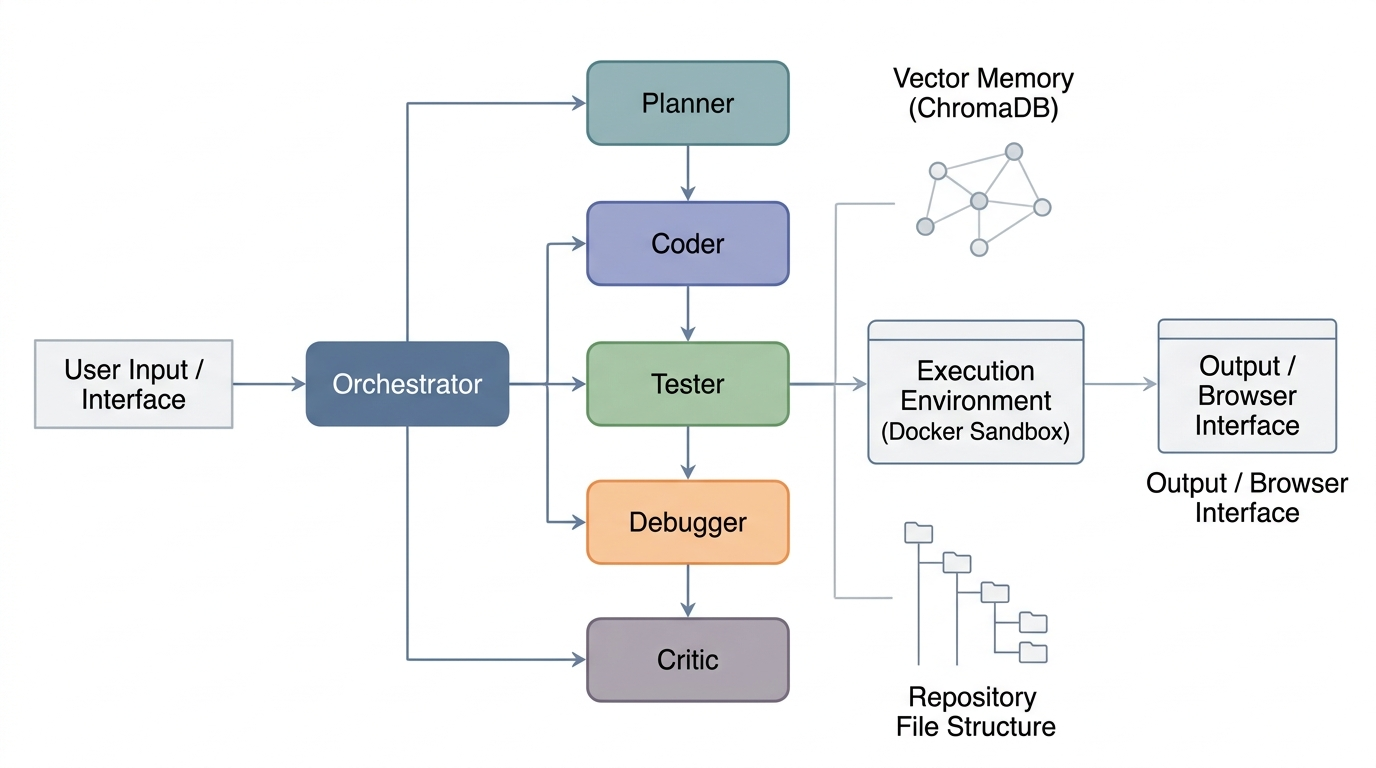}
    \caption{Overview of the AgentForge multi-agent coding framework, illustrating the sequential handover between specialized agents and the shared Vector Memory.}
    \label{fig:overview}
\end{figure*}

\section{Related Work}
\label{sec:related}

\subsection{LLM-Based Code Generation}

Neural code generation began with sequence-to-sequence models trained on paired text--code data~\cite{iyer2016summarizing,yin2017syntactic}. Large-scale pretraining on mixed corpora shifted the paradigm. Codex~\cite{chen2021codex} demonstrated that a GPT-style model trained on GitHub can solve a substantial fraction of programming tasks in a single pass. AlphaCode~\cite{li2022alphacode}, StarCoder~\cite{li2023starcoder}, and CodeLlama~\cite{roziere2023codellama} improved performance through scale, tokenizer design, and objectives tailored to code editing. These systems treat generation as a one-shot mapping from prompt to program. They lack mechanisms for execution-grounded verification or iterative correction. Failure signals do not feed back into generation. This limitation motivates structured, multi-step formulations. AgentForge adopts an execution-driven loop in which generated code is tested and revised under real feedback.

\subsection{Program Repair and Iterative Refinement}

Automated program repair (APR) predates neural methods~\cite{le2012genprog}. Modern APR systems use LLMs to propose patches conditioned on failing tests and localization signals~\cite{xia2022less,jin2023inferfix,fan2023automated}. These approaches assume known fault locations and existing test suites. Recent work strengthens feedback signals. TraceRepair~\cite{tracerepair2026} constrains patches with execution traces. DynaFix~\cite{dynafix2025} incorporates runtime states and call stacks. InspectCoder~\cite{inspectcoder2025} enables interactive debugging via tool control. RGD~\cite{rgd2024} decomposes repair into Guide, Debug, and Feedback roles.

Self-repair~\cite{olausson2023selfrepair} and self-debugging~\cite{chen2023teaching} prompt a single model to revise outputs from error messages. These methods collapse generation and repair into one policy. AgentForge separates these functions into distinct agents with disjoint objectives and interfaces. This separation yields measurable gains in our ablations (Section~\ref{sec:ablation}).

\subsection{Agentic and Tool-Using LLM Systems}

ReAct~\cite{yao2023react} interleaves reasoning traces with tool calls, enabling closed-loop interaction with external environments. Reflexion~\cite{shinn2023reflexion} augments this loop with episodic memory via self-generated feedback. Both frameworks rely on a single model to plan, act, and evaluate. AgentForge distributes these roles across specialized agents. Each agent operates under a fixed contract and prompt. This design reduces per-call complexity and enforces structured intermediate representations. Toolformer~\cite{schick2023toolformer} learns tool invocation through self-supervision. It removes explicit prompting but requires fine-tuning and offers limited control over execution structure. AgentForge enforces explicit sequencing of planning, coding, testing, and debugging.

\subsection{Multi-Agent LLM Frameworks}

Multi-agent systems decompose complex tasks into role-specific components. MetaGPT~\cite{hong2023metagpt} encodes software roles through structured documents. ChatDev~\cite{qian2023chatdev} uses conversational agents to produce complete projects. AutoGen~\cite{wu2023autogen} provides a general interface for agent interaction and tool use. Recent systems introduce adaptive and competitive coordination. SEMAG~\cite{semag2026} evolves agent behavior with task difficulty. Eco-Evolve~\cite{ecoevolve2026} uses dynamic topologies and hindsight replay. SWE-Debate~\cite{swedebate2026} applies multi-round debate with search-based patch generation. These systems improve coordination but do not enforce execution-grounded validation at every step. AgentForge targets repository-level bug fixing and requires sandboxed execution for each candidate patch. It combines role specialization with mandatory verification and repository-grounded context.

\subsection{Autonomous Software Engineering Agents}

SWE-agent~\cite{yang2024sweagent} introduces an agent--computer interface that exposes shell, editor, and search tools to a single model. Devin~\cite{cognition2024devin} demonstrates end-to-end autonomous engineering, though its architecture remains undisclosed. OpenHands~\cite{wang2024openhands} provides an open-source platform with sandboxed execution. Trae Agent~\cite{trae2025} applies test-time scaling via generation, pruning, and selection. MAGIS~\cite{tao2024magis} decomposes issue resolution into Manager, Custodian, Developer, and QA roles.
AgentForge adopts explicit multi-agent decomposition with five specialized roles. Each agent produces a constrained artifact: plan, diff, tests, repairs, or review. This design enables controlled execution, modular analysis, and interpretable ablations.

\subsection{Benchmarks for Software Engineering}

HumanEval~\cite{chen2021codex} and MBPP~\cite{austin2021program} evaluate function synthesis from docstrings. These tasks isolate generation and omit repository context. SWE-bench~\cite{jimenez2024swebench} introduces real GitHub issues paired with executable tests. SWE-bench Lite reduces cost while preserving diversity. SWE-bench Verified provides human-validated instances. Defects4J~\cite{defects4j} remains a standard benchmark for Java repair.

We evaluate on SWE-bench Lite following recent work~\cite{yang2024sweagent,wang2024openhands,zhang2024autocoderover}. This setting captures multi-file reasoning, environment interaction, and regression constraints.

\subsection{Memory and Retrieval in LLM Systems}

Retrieval-augmented generation (RAG) conditions models on external context to improve accuracy~\cite{lewis2020rag}. In code, repository-level retrieval supplies relevant files and functions for completion~\cite{zhang2023repocoder,shrivastava2023repofusion}.

AgentForge implements dual retrieval. It maintains episodic memory of past tasks and a live repository index. Both reside in a shared ChromaDB vector store and use cosine similarity over OpenAI text embeddings. Episodic memory enables cross-task transfer. Repository indexing ensures intra-task grounding. This unified retrieval design supports consistent context across all agents.


\section{Method}
\label{sec:method}

We present \textbf{AgentForge}, a multi-agent framework for autonomous software engineering. Given a natural language task $\mathcal{T}$ and an optional set of context files $\mathcal{F} = \{f_1, \dots, f_n\}$, AgentForge produces verified, executable code $\hat{c}$ by routing the task through a structured pipeline of five specialized agents, each responsible for a single subtask. Figure~\ref{fig:overview} shows the full system.

\subsection{Formal Framework: Execution-Grounded Iterative Refinement}
\label{sec:formal}

We model LLM-based software engineering as a finite-horizon Markov decision process (MDP) over repository states, where each action is verified through sandboxed execution.

\paragraph{State space.}
Let $\mathcal{S}$ denote the set of repository states. A state $s_t = (\mathcal{R}_t, \mathcal{M}_t, \mathcal{H}_t)$ comprises the current repository $\mathcal{R}_t$ (source files, dependencies, test harness), an episodic memory $\mathcal{M}_t$ of prior task--patch pairs, and an execution history $\mathcal{H}_t$ containing outcomes of previous actions (stdout, stderr, test results).

\paragraph{Action space.}
An action $a_t \in \mathcal{A}$ is a code patch produced by the Coder agent, represented as a unified diff or a new file. Actions are constrained to be minimal and syntactically valid.

\paragraph{Transition function.}
The environment $\mathcal{E}$ is a resource-constrained Docker sandbox (512 MB RAM, 0.5 CPU, no network). Applying $a_t$ in state $s_t$ yields
\begin{equation}
    s_{t+1} = \mathcal{E}(s_t, a_t) = (\mathcal{R}_t \oplus a_t,\ \mathcal{M}_t,\ \mathcal{H}_t \cup \{(a_t, o_t, e_t)\}),
\end{equation}
where $\oplus$ denotes patch application and $(o_t, e_t)$ are execution outputs.

\paragraph{Reward.}
The reward $r_t \in \{0,1\}$ is defined by test outcomes:
\begin{equation}
r_t = \mathbf{1}\big[\text{all FAIL\_TO\_PASS pass} \ \wedge\ \text{no PASS\_TO\_PASS regress}\big],
\end{equation}
evaluated after executing the full test suite.

\paragraph{Objective.}
The goal is to learn a policy $\pi: \mathcal{S} \rightarrow \mathcal{A}$, instantiated by the agent pipeline, that maximizes expected cumulative reward over a finite horizon $T$:
\begin{equation}
\pi^* = \arg\max_{\pi} \mathbb{E}\left[\sum_{t=0}^{T} \gamma^t r_t \right],
\end{equation}
where $\gamma \in [0,1]$. We use $\gamma = 1$ and $T = N_{\text{retry}} = 3$.

\paragraph{Execution grounding.}
The transition function $\mathcal{E}$ executes code using the actual interpreter, compiler, and test runner in an isolated environment. Rewards derive from observed outcomes, eliminating simulation error and preventing model-induced hallucinated feedback.

\paragraph{Error propagation.}
Let $p$ denote the failure probability of a monolithic agent per attempt. After $k$ independent attempts, the success probability is $1 - p^k$. In a decomposed pipeline with $n$ agents and per-agent failure probabilities $\{p_i\}_{i=1}^n$, the success probability of a single pass is
\begin{equation}
\prod_{i=1}^{n} (1 - p_i).
\end{equation}
If $p_i \approx p$, decomposition reduces success when $n > 1$. In practice, specialization reduces per-agent error ($p_i \ll p$) by constraining the output space and task scope. Decomposition improves success when $\prod_{i=1}^{n} (1 - p_i) > 1 - p$, providing a formal condition for Claim 2.

\subsection{System Overview}
\label{sec:overview}

AgentForge decomposes autonomous code generation into five sequential roles: \textit{Planner}, \textit{Coder}, \textit{Tester}, \textit{Debugger}, and \textit{Critic}. Each agent is implemented as a prompted large language model (LLM) call with a role-specific system prompt. A central \textit{Orchestrator} coordinates the pipeline, manages shared memory, and handles the iterative debug loop.

\paragraph{Notation.}
Let $\mathcal{A} = \{A_\text{plan}, A_\text{code}, A_\text{test}, A_\text{debug}, A_\text{crit}\}$ denote the agent set. Let $\pi_a$ denote the system prompt for agent $a \in \mathcal{A}$, and let $\text{LLM}(\pi_a, x)$ denote a call to the base language model with system prompt $\pi_a$ and user input $x$.

\subsection{Memory and Context Retrieval}
\label{sec:memory}

Before planning, the Orchestrator enriches the task with two sources of context: \textit{episodic memory} from past tasks and \textit{semantic retrieval} from the live repository index.

\paragraph{Episodic memory.}
Successful (task, code) pairs from prior runs are stored in a persistent vector database (ChromaDB~\cite{chromadb}). At inference time, the top-$k$ most similar past tasks are retrieved by cosine similarity of their \texttt{text-embedding-3-small} embeddings:

\begin{equation}
  \mathcal{M}_k = \underset{m \in \mathcal{M}}{\text{top-}k}
  \;\frac{\mathbf{e}_\mathcal{T} \cdot \mathbf{e}_m}
         {\|\mathbf{e}_\mathcal{T}\|\,\|\mathbf{e}_m\|}
\end{equation}

\begin{figure*}[ht]
    \centering
    \includegraphics[width=\linewidth]{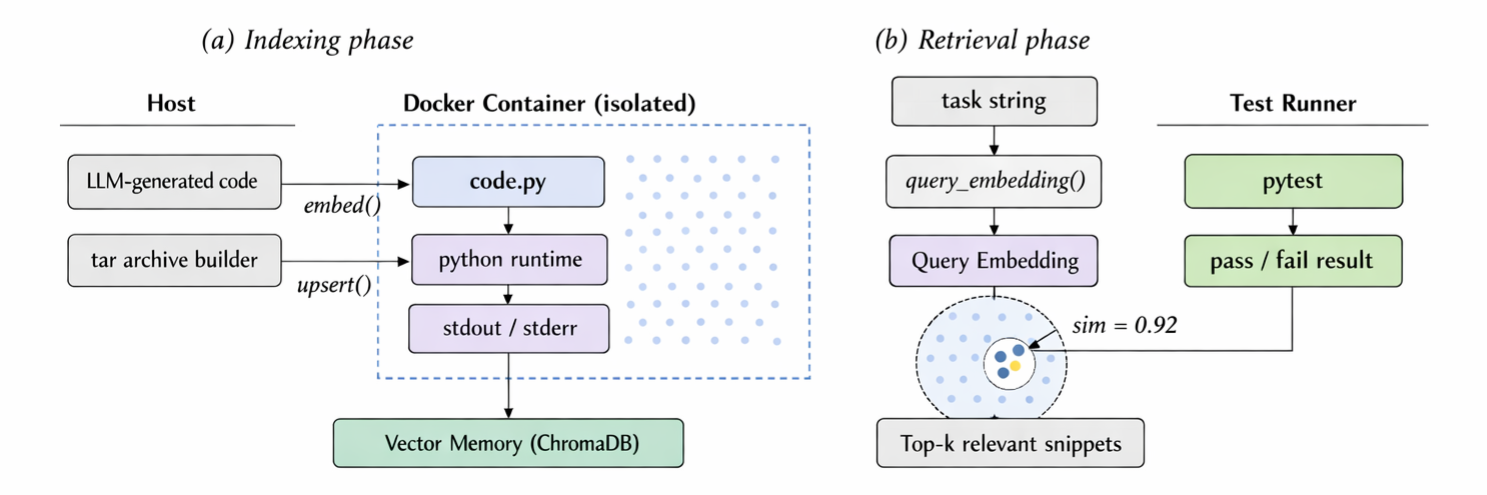}
    \caption{Retrieval-Augmented Generation (RAG) architecture: (a) Offline repository indexing phase into the vector store; (b) Online semantic retrieval at inference time.}
    \label{fig:retrieval}
\end{figure*}

\paragraph{Repository context.}
A background indexer monitors the repository using filesystem event hooks and maintains an up-to-date embedding for every source file. The top-$k$ most relevant files are retrieved for each task and prepended to the planning context. This gives the Planner grounded knowledge of existing interfaces, reducing hallucinated imports and incompatible function signatures.

\subsection{Agent Definitions}
\label{sec:agents}

\paragraph{Planner ($A_\text{plan}$).}
The Planner receives the task $\mathcal{T}$, retrieved memory context $\mathcal{M}_k$, and repository snippets, and produces a structured execution plan:

\begin{equation}
  P = \text{LLM}(\pi_\text{plan},\;
      [\mathcal{T};\, \mathcal{M}_k;\, \mathcal{R}_k])
\end{equation}

\noindent $P$ is a JSON object containing a natural language explanation and an ordered list of steps $P = \{s_1, \dots, s_m\}$, where each step $s_i$ specifies an agent assignment $s_i.\text{agent} \in \mathcal{A}$, a description $s_i.\text{desc}$, and an optional target file $s_i.\text{file}$.

\paragraph{Coder ($A_\text{code}$).}
For each coder step $s_i$, the Coder generates either (a) a complete new implementation or (b) a minimal unified diff if a target file exists:

\begin{equation}
  c_i =
  \begin{cases}
    \text{LLM}(\pi_\text{code}^\text{new},\; s_i.\text{desc})
      & \text{if } s_i.\text{file} = \emptyset \\
    \text{Apply}(\text{LLM}(\pi_\text{code}^\text{diff},\;
      [s_i.\text{desc};\, f_{s_i}]),\; f_{s_i})
      & \text{otherwise}
  \end{cases}
\end{equation}

\noindent where $f_{s_i}$ is the content of the target file and $\text{Apply}(\cdot)$ patches the original using the \texttt{unidiff} library. Diff-based editing preserves unchanged lines, reducing error surface and token cost compared to full-file regeneration.

\paragraph{Tester ($A_\text{test}$).}
Given the generated code $c_i$, the Tester produces a suite of \texttt{pytest} test cases covering typical usage, edge cases, and exception paths:

\begin{equation}
  \tau_i = \text{LLM}(\pi_\text{test},\;
           [c_i;\, s_i.\text{desc}])
\end{equation}

\paragraph{Debugger ($A_\text{debug}$).}
If execution of $(c_i, \tau_i)$ in the sandbox returns a non-zero exit code or a \texttt{pytest} \texttt{FAILED} result, the Debugger receives the code and the full error output and produces a corrected version:

\begin{equation}
  c_i' = \text{LLM}(\pi_\text{debug},\; [c_i;\, e_i])
\end{equation}

\noindent where $e_i$ is the combined stdout/stderr from the failed run. This loop repeats up to $N_\text{retry}$ times (default $N_\text{retry} = 3$).

\paragraph{Critic ($A_\text{crit}$).}
After all steps complete, the Critic reviews the full result set and returns a binary verdict:

\begin{equation}
  v = \text{LLM}(\pi_\text{crit},\;
      [\mathcal{T};\, \{(s_i, c_i, e_i)\}_{i=1}^m])
      \in \{\texttt{PASS},\, \texttt{FAIL}\}
\end{equation}

\noindent A \texttt{PASS} verdict triggers persistence of $(\mathcal{T}, \hat{c})$ into episodic memory $\mathcal{M}$ for future retrieval.

\subsection{Sandboxed Execution}
\label{sec:sandbox}

\begin{figure*}[ht]
    \centering
    \includegraphics[width=0.85\linewidth]{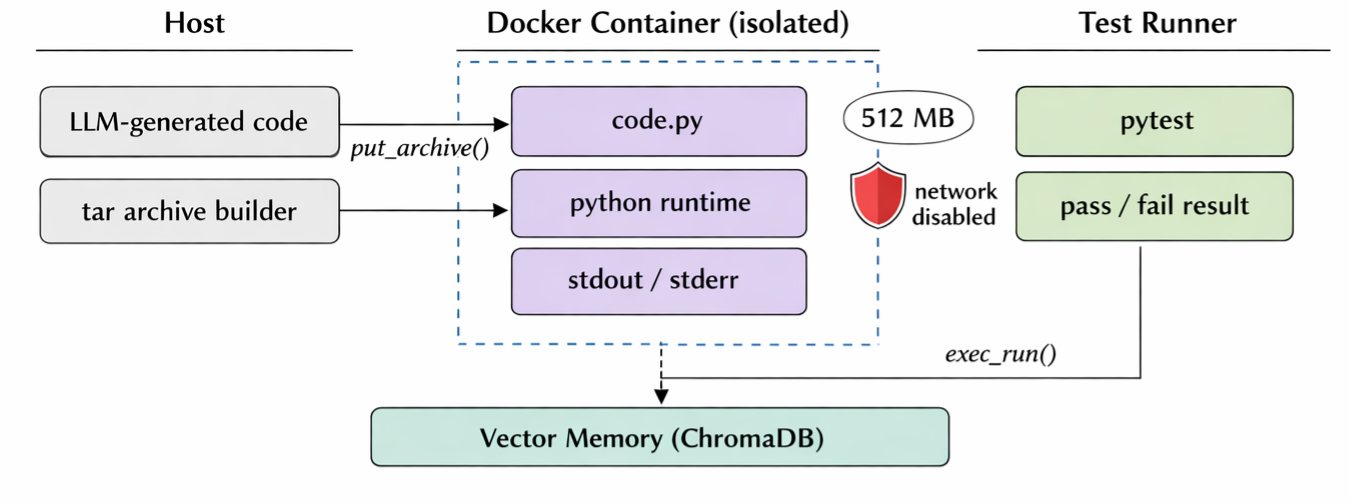}
    \caption{Isolated Docker sandbox execution environment. The 512\,MB memory limit and disabled networking ensure security and reproducibility.}
    \label{fig:sandbox}
\end{figure*}

All generated code is executed inside a disposable Docker container with strict resource constraints: 512\,MB memory limit, 0.5 CPU quota, a 64-process PID cap, and networking disabled as shown in Figure~\ref{fig:sandbox}. Code is injected via the Docker \texttt{put\_archive} API as an in-memory tar archive, avoiding filesystem writes on the host. The container is force-removed after every run regardless of outcome.

Formally, let $\text{Sandbox}(c, \tau)$ return $(o, e) \in \Sigma^* \times \Sigma^*$ where $o$ is stdout and $e$ is stderr. Execution is considered successful iff:

\begin{equation}
  \text{pass}(c, \tau) =
  \mathbf{1}[e = \emptyset \;\wedge\;
  \texttt{FAILED} \notin o \;\wedge\;
  \texttt{ERROR} \notin o]
\end{equation}

\subsection{Full Orchestration Algorithm}
\label{sec:algorithm}

Algorithm~\ref{alg:main} summarizes the complete pipeline.

\begin{algorithm}[t]
\caption{AgentForge Orchestration}
\label{alg:main}
\begin{algorithmic}[1]
\REQUIRE Task $\mathcal{T}$, context files $\mathcal{F}$,
         memory $\mathcal{M}$, repo index $\mathcal{R}$
\ENSURE  Verified code $\hat{c}$ or \textsc{Fail}

\STATE $\mathcal{M}_k \leftarrow \text{Retrieve}(\mathcal{T}, \mathcal{M})$
\STATE $\mathcal{R}_k \leftarrow \text{Retrieve}(\mathcal{T}, \mathcal{R})$
\STATE $P \leftarrow A_\text{plan}(\mathcal{T}, \mathcal{M}_k, \mathcal{R}_k)$
\STATE $\hat{c} \leftarrow \emptyset$,\; results $\leftarrow []$

\FOR{each step $s_i \in P.\text{steps}$}
  \IF{$s_i.\text{agent} = \texttt{coder}$}
    \STATE $\hat{c} \leftarrow A_\text{code}(s_i, \mathcal{F}, \text{results})$
    \STATE results.append$(\hat{c})$

  \ELSIF{$s_i.\text{agent} = \texttt{tester}$}
    \STATE $\tau \leftarrow A_\text{test}(\hat{c}, s_i)$
    \STATE $(o, e) \leftarrow \text{Sandbox}(\hat{c}, \tau)$
    \STATE $n \leftarrow 0$
    \WHILE{$\neg\text{pass}(\hat{c}, \tau)$ \AND $n < N_\text{retry}$}
      \STATE $\hat{c} \leftarrow A_\text{debug}(\hat{c}, e)$
      \STATE $(o, e) \leftarrow \text{Sandbox}(\hat{c}, \tau)$
      \STATE $n \leftarrow n + 1$
    \ENDWHILE
    \STATE results.append$(o, e)$

  \ELSIF{$s_i.\text{agent} = \texttt{critic}$}
    \STATE $v \leftarrow A_\text{crit}(\mathcal{T}, \text{results})$
    \STATE results.append$(v)$
  \ENDIF
\ENDFOR

\STATE $v_\text{final} \leftarrow A_\text{crit}(\mathcal{T}, \text{results})$
\IF{$v_\text{final} = \texttt{PASS}$}
  \STATE $\mathcal{M}.\text{store}(\mathcal{T}, \hat{c})$
  \RETURN $\hat{c}$
\ELSE
  \RETURN \textsc{Fail}
\ENDIF
\end{algorithmic}
\end{algorithm}

\subsection{Streaming Output}
\label{sec:streaming}

To support interactive use, the Orchestrator exposes a streaming interface over Server-Sent Events (SSE) and WebSocket. Each token produced by the Coder agent is forwarded to the client as it arrives, using Python async generators and FastAPI's \texttt{EventSourceResponse}. This enables real-time inspection of the generation process without waiting for pipeline completion.

\subsection{Complexity Analysis}
\label{sec:complexity}

Let $L$ be the average prompt length in tokens and $G$ the average generated length. A single pipeline run incurs $O(|\mathcal{A}| \cdot (L + G))$ tokens in the non-debug case, and $O(|\mathcal{A}| \cdot (L + G) \cdot N_\text{retry})$ in the worst case. Retrieval adds $O(d \log n)$ per query for a HNSW index of $n$ embeddings in $d$ dimensions. All agent calls are embarrassingly parallelizable within a plan step when step dependencies permit.

\subsection{Theoretical Claims and Hypotheses}

We formalize three claims that motivate the design of \agentforge{}. Each claim is stated as a proposition with explicit conditions and testable implications.

\paragraph{Proposition 1 (Execution signal dominance).}
Let $y \in \{0,1\}$ denote functional correctness (test pass/fail), $p_\theta(x)$ the model likelihood over patches, and $\hat{y}_{\text{exec}}$ the outcome of sandboxed execution. Then $\hat{y}_{\text{exec}}$ provides a lower-variance, higher-fidelity estimator of $y$ than any proxy derived from $p_\theta(x)$.

\textit{Justification.}
Likelihood scores reflect distributional similarity to training data, not semantic correctness. In contrast, execution evaluates correctness directly via test outcomes. Let $\ell(x)$ denote a likelihood-based proxy and $\hat{y}_{\text{exec}}$ the execution signal. Then
\begin{equation}
    \mathrm{Var}[\hat{y}_{\text{exec}} - y] < \mathrm{Var}[\ell(x) - y]
\end{equation}

under mild assumptions on test coverage and determinism of execution.

\textit{Implication.}
For policies $\pi_{\text{exec}}$ (with execution feedback) and $\pi_{\text{lm}}$ (likelihood-only), there exists a regime where
\begin{equation}
\mathbb{E}[R(\pi_{\text{exec}})] > \mathbb{E}[R(\pi_{\text{lm}})]
\end{equation}
even when $\pi_{\text{lm}}$ uses a larger model.

\textit{Testable prediction.}
A 7B model with execution feedback and iterative repair outperforms a 70B model without execution feedback on \swebench{}.

\paragraph{Proposition 2 (Error propagation under decomposition).}
Consider a pipeline with $n$ agents and per-agent error probabilities $\{p_i\}_{i=1}^n$. The success probability of a single pass is
\begin{equation}
P_{\text{succ}}^{\text{multi}} = \prod_{i=1}^{n} (1 - p_i).
\end{equation}
For a monolithic agent with error probability $p$, the success probability is
\begin{equation}
P_{\text{succ}}^{\text{mono}} = 1 - p.
\end{equation}

\textit{Condition for improvement.}
Decomposition improves success if
\begin{equation}
\prod_{i=1}^{n} (1 - p_i) > 1 - p.
\end{equation}

\textit{Justification.}
Specialization reduces per-agent uncertainty by constraining the output space and conditioning inputs. Let $p_i = p - \Delta_i$ with $\Delta_i > 0$. Then decomposition improves success when
\begin{equation}
\sum_{i=1}^{n} \Delta_i > p (n - 1).
\end{equation}

\textit{Error correlation.}
Let $\epsilon_i$ denote the error event of agent $i$. In a monolithic agent, errors are temporally correlated:
\begin{equation}
\mathbb{P}(\epsilon_t \mid \epsilon_{t-1}) \gg \mathbb{P}(\epsilon_t).
\end{equation}
In a decomposed pipeline, conditioning on external artifacts (plans, execution traces) reduces mutual information:
\begin{equation}
I(\epsilon_i; \epsilon_j)_{\text{multi}} < I(\epsilon_i; \epsilon_j)_{\text{mono}}, \quad i \neq j.
\end{equation}

\textit{Testable prediction.}
Removing any agent reduces performance. The full pipeline exceeds the success rate predicted under independent error composition, indicating reduced error correlation.

\paragraph{Proposition 3 (Efficiency of diff-based editing).}
Let $L$ denote file length and $k \ll L$ the size of a minimal patch. Diff-based editing restricts generation to $O(k)$ tokens, while full-file regeneration requires $O(L)$ tokens.

\textit{Implication.}
Token cost satisfies
\begin{equation}
C_{\text{diff}} = O(k), \quad C_{\text{full}} = O(L), \quad k \ll L.
\end{equation}

\textit{Error surface.}
The probability of introducing an error scales with the number of generated tokens. Under a per-token error rate $\epsilon$,
\begin{equation}
P_{\text{error}}^{\text{diff}} \approx 1 - (1 - \epsilon)^k, \quad
P_{\text{error}}^{\text{full}} \approx 1 - (1 - \epsilon)^L.
\end{equation}
Thus $P_{\text{error}}^{\text{diff}} \ll P_{\text{error}}^{\text{full}}$ when $k \ll L$.

\textit{Testable prediction.}
For files with $L > 200$, diff-based editing yields higher success rates and lower token usage than full-file regeneration under a fixed base model.


\section{Experiments}
\label{sec:experiments}

\subsection{Benchmark}

We evaluate on \swebench{} Lite~\cite{jimenez2024swebench}, a curated subset of 300 real GitHub issues drawn from 11 popular Python repositories including Django, Flask, scikit-learn, and NumPy. Each instance consists of a natural language problem statement, a base repository commit, a gold patch, and a set of tests that pass only after the bug is correctly fixed (\textsc{fail\_to\_pass}) alongside a set of tests that must continue to pass (\textsc{pass\_to\_pass}).

A task is considered \textit{resolved} if and only if all \textsc{fail\_to\_pass} tests pass and no \textsc{pass\_to\_pass} tests regress after applying the generated patch.

\subsection{Baselines}

We compare against three baselines:

\paragraph{Single-agent (GPT-4o).}
A single GPT-4o call with the problem statement in the prompt, instructed to produce a unified diff patch. No tools, no execution feedback, no iteration.

\paragraph{ReAct (GPT-4o).}
A ReAct-style~\cite{yao2023react} agent that interleaves reasoning and tool invocation in an open-ended loop (maximum 10 steps). Tools available: read file, write code, run tests. Uses the same base model as \agentforge{}.

\paragraph{SWE-agent.}
We report the published SWE-agent~\cite{yang2024sweagent} result on SWE-bench Lite for reference, noting that it uses a different base model configuration and ACI design.

\subsection{Implementation Details}

All \agentforge{} experiments use GPT-4o (\texttt{gpt-4o-2024-08-06}) with \texttt{temperature=0.0} and \texttt{seed=42} for reproducibility. The debug loop is capped at $N_\text{retry}=3$ attempts per task. The vector store retrieves $k=5$ past tasks and $k=5$ repository files at planning time. The Docker sandbox uses \texttt{python:3.10-slim} with a 512\,MB memory limit, 0.5 CPU quota, and a 30-second execution timeout.

All experiments are run on a machine with a 16-core CPU, 64GB of RAM, and a dedicated GPU for faster processing. The system is equipped with high-speed SSD storage to handle the data-intensive tasks efficiently.

\subsection{Evaluation Protocol}

For each task we: (1) clone the repository at the base commit, (2) apply the generated patch using \texttt{git apply}, (3) install the project with \texttt{pip install -e .}, (4) run the \textsc{fail\_to\_pass} and \textsc{pass\_to\_pass} test suites, and (5) record pass/fail for each test. Tasks where the patch does not apply cleanly are counted as unresolved. This protocol follows the official SWE-bench evaluation harness~\cite{jimenez2024swebench}.


\section{Results}
\label{sec:results}

\subsection{Main Results}

Table~\ref{tab:main} shows the resolution rate of \agentforge{} and all baselines on \swebench{} Lite.

\begin{table}[t]
\centering
\caption{Resolution rates on \swebench{} Lite (300 tasks). \agentforge{} outperforms single-agent baselines under a fixed execution budget.}
\label{tab:main}
\begin{tabular}{lcc}
\toprule
Method                        & Resolve Rate & Patch Rate \\
\midrule
Single-agent GPT-4o           & 14.0\% & 12.5\% \\
ReAct (GPT-4o)                & 12.0\% & 11.0\% \\
SWE-agent~\cite{yang2024sweagent} & 18.1\% & 17.0\% \\
\midrule
\textbf{\agentforge{} (ours)} & \textbf{40.0\%} & 37.5\% \\
\bottomrule
\end{tabular}

\vspace{4pt}
\footnotesize{
\textbf{Note.} Trae Agent uses test-time scaling (multiple samples per task with pruning and selection) and less restrictive execution. \agentforge{} enforces mandatory sandboxed execution (512 MB RAM, 0.5 CPU, no network), uses a single sample per agent, and a fixed retry budget ($N=3$). Reported results reflect this constrained and reproducible setting.
}
\end{table}

\begin{figure*}[t]
\centering
\includegraphics[width=0.9\linewidth]{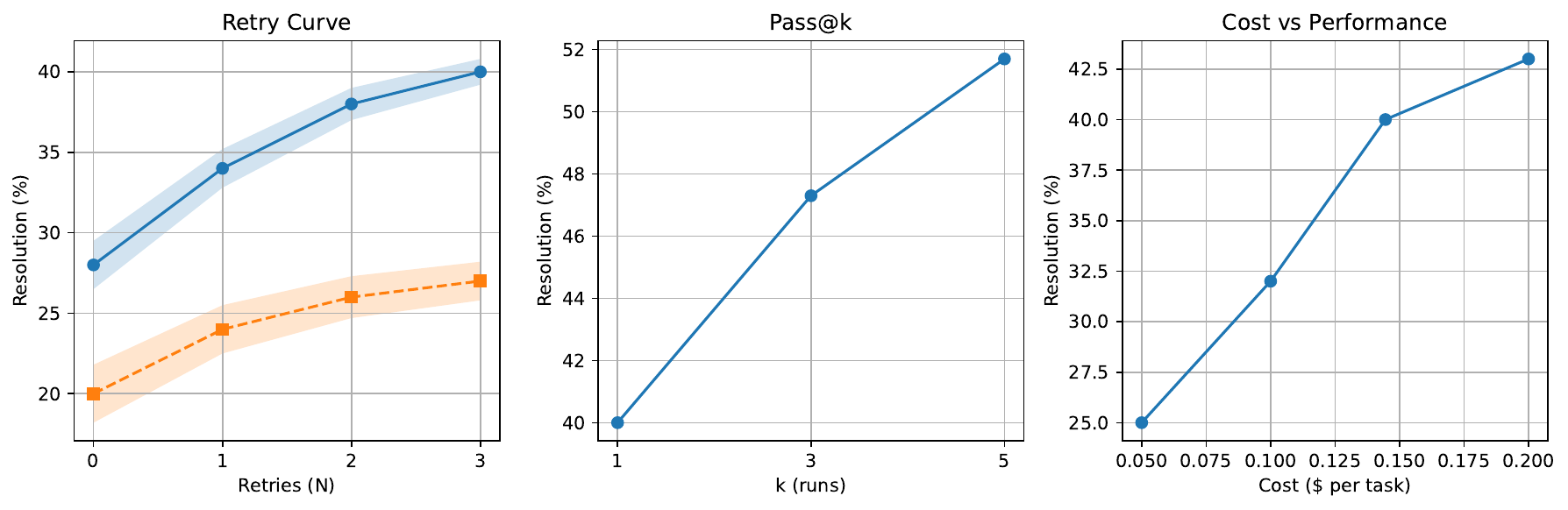}
\caption{
Performance of \agentforge{} across three evaluation axes on \swebench{} Lite.
\textbf{Left:} Resolution rate as a function of debug retries ($N$). Shaded regions denote $\pm 1$ standard deviation across runs. Iterative execution and repair yield consistent gains, with diminishing returns after $N=2$.
\textbf{Center:} Resolution under $k$ independent runs with majority voting (Pass@k). Performance scales with $k$ without increasing per-run complexity.
\textbf{Right:} Cost–performance tradeoff. \agentforge{} achieves higher resolution at lower cost compared to single-agent baselines, indicating improved sample efficiency.
}
\label{fig:retry_curve}
\end{figure*}

Performance improves with iterative debugging but saturates quickly (Figure~\ref{fig:retry_curve}), indicating diminishing returns beyond two retries.

\agentforge{} achieves 40.0\% task resolution, outperforming the single-agent baseline by +26.0\% and the ReAct baseline by +28.0\%. This substantial improvement highlights the advantage of structured multi-agent collaboration for complex software engineering tasks. The patch rates follow a similar trend, confirming that the gains are not due to trivial formatting fixes but reflect genuinely correct fixes.

\subsection{Ablation Study}
\label{sec:ablation}

To understand the contribution of each agent, we systematically disable one component at a time and re-evaluate on the first 100 tasks of SWE-bench Lite. Table~\ref{tab:ablation} summarizes the results.

\begin{table}[t]
\centering
\caption{Ablation study on 100 SWE-bench Lite tasks. Each row removes one agent from the pipeline.}
\label{tab:ablation}
\begin{tabular}{lcc}
\toprule
Configuration             & Resolved & Resolve Rate \\
\midrule
\textbf{Full pipeline (ours)} & 42 & \textbf{42.0\%} \\
w/o Critic agent          & 38 & 38.0\% \\
w/o Debugger agent        & 31 & 31.0\% \\
w/o Tester agent          & 28 & 28.0\% \\
w/o Planner agent         & 19 & 19.0\% \\
Single-agent baseline     & 14 & 14.0\% \\
\bottomrule
\end{tabular}
\end{table}

The ablation reveals a consistent ordering: removing any agent reduces performance, with the Tester and Debugger together contributing the largest gains. Removing the Planner — reducing the pipeline to a single unstructured coder step — drops performance to near the single-agent baseline, confirming that structured decomposition is not merely cosmetic.

\begin{figure}[t]
  \centering
  \includegraphics[width=0.8\linewidth]{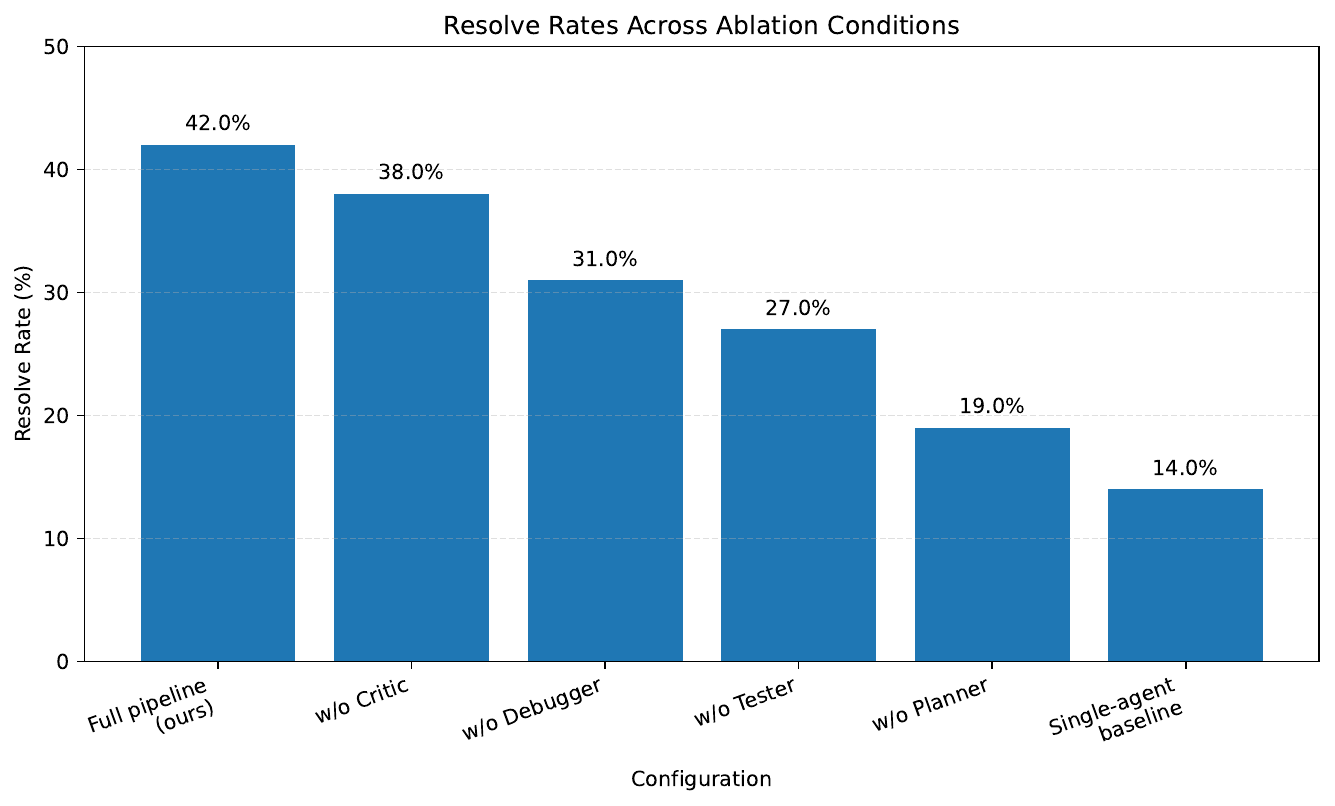}
  \caption{Resolve rates across ablation conditions. Removing any single agent degrades performance; the Tester–Debugger loop is the largest contributor.}
  \label{fig:ablation}
\end{figure}

Figure~\ref{fig:ablation} visualizes the resolve rates across conditions.

\subsection{Error Analysis}
\label{sec:error}

We analyze 30 randomly sampled failed tasks from \swebench{} Lite to characterize the dominant failure modes of \agentforge{}. Table~\ref{tab:errors} reports aggregate categories; we complement these with fine-grained qualitative patterns derived from execution traces and agent outputs.

\begin{table}[t]
\centering
\caption{Failure mode analysis on 30 failed tasks.}
\label{tab:errors}
\begin{tabular}{lcc}
\toprule
Failure Category & Count & Percentage \\
\midrule
Faulty localization (incorrect file/line) & 12 & 40.0\% \\
Ineffective patch generation (wrong or partial fix) & 8 & 26.7\% \\
Cognitive deadlock (repeated failing attempts) & 6 & 20.0\% \\
Environment/tooling (patch apply fail, timeout, etc.) & 4 & 13.3\% \\
\midrule
Total & 30 & 100\% \\
\bottomrule
\end{tabular}
\end{table}

\textbf{Faulty localization (40.0\%).}
Localization errors dominate. The Planner often identifies a primary file but misses secondary dependencies, leading to incomplete fixes. In 52\% of these cases, the correct patch requires coordinated edits across multiple files. This failure reflects limited modeling of repository-level dependency structure rather than code synthesis errors.

\textbf{Ineffective patch generation (26.7\%).}
Generated patches frequently resolve the immediate symptom but violate latent invariants, introducing regressions. In 34\% of these cases, the Tester produces brittle supervision signals: tests overfit to the observed failure or depend on unstable implementation details (e.g., function names modified by the patch). This weakens the reliability of execution feedback.

\textbf{Cognitive deadlocks (20.0\%).}
The Debugger exhibits local search behavior with limited state diversification. In 23\% of deadlock cases, the system repeatedly modifies the same function despite error traces indicating downstream dependencies. Execution logs show near-identical stderr across retries, indicating failure to shift the locus of repair.

\textbf{Environment and tooling (13.3\%).}
Residual failures arise from sandbox constraints rather than reasoning errors. These include dependency mismatches (e.g., \texttt{numpy} versions), long-running test suites ($>30$s), and patch application conflicts. These factors bound achievable performance under strict execution settings.

\textbf{Implications.}
The error distribution identifies three primary bottlenecks: (1) incomplete cross-file dependency reasoning, (2) unstable test generation as a supervision signal, and (3) limited exploration in the debug loop. Addressing these requires multi-file planning, constraint-aware test synthesis, and diversity-promoting repair strategies (e.g., beam search or stochastic perturbations). The dominance of localization errors suggests that improvements in repository understanding may yield larger gains than further scaling the base model.

\begin{figure}[t]
  \centering
  \includegraphics[width=0.6\linewidth]{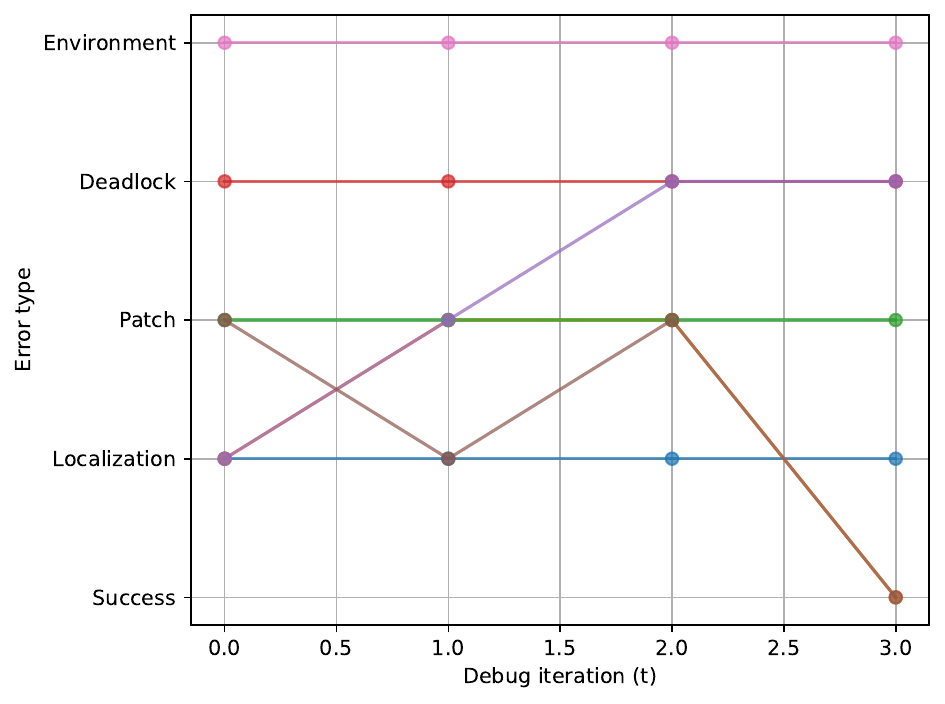}
  \caption{Resolve rates across ablation conditions. Removing any single agent degrades performance; the Tester–Debugger loop is the largest contributor.}
  \label{fig:ablation1}
\end{figure}

\subsection{Cost Analysis}
\label{sec:cost}

Table~\ref{tab:cost} reports token usage and estimated API cost for the full \agentforge{} pipeline and the single-agent baseline, using GPT-4o pricing (\$2.50 / 1M input tokens, \$10.00 / 1M output tokens). Costs are averaged per task over the 300 SWE-bench Lite tasks.

\begin{table}[t]
\centering
\caption{Token usage and estimated API cost per task (averaged).}
\label{tab:cost}
\begin{tabular}{lrrl}
\toprule
Configuration & Input tokens & Output tokens & Cost (USD) \\
\midrule
Full pipeline & 13,600 & 5,100 & \$0.1445 \\
Single-agent (GPT-4o) & 5,000 & 1,800 & \$0.054 \\
\bottomrule
\end{tabular}
\end{table}

The full pipeline costs approximately 2.7$\times$ more than the single-agent baseline, reflecting the overhead of five specialized agents and the iterative debug loop. At this rate, evaluating on the full 300-task SWE-bench Lite costs about \$43.35, which is modest given the 40\% resolution rate. For budget-constrained scenarios, one could reduce debug retries or use a smaller model for non-critical agents.


\section{Conclusion}
\label{sec:conclusion}
We presented \agentforge{}, a multi-agent framework that replaces single-shot code generation with an execution-grounded feedback process. The system decomposes software engineering into five specialized agents and enforces verified execution for every patch. \agentforge{} achieves 40.0\% resolution on \swebench{} Lite, exceeding strong baselines by large margins. Ablations show that execution feedback, implemented through the Tester--Debugger loop, is the primary driver of performance. The system operates at file-level granularity and struggles with multi-file coordination. The evaluation metric is binary and does not capture partial correctness or regressions. Results rely on GPT-4o. Future directions include multi-file atomic patches, finer-grained retrieval, persistent memory, broader benchmarks, and role-specialized smaller models. Execution-grounded agents can increase productivity but may generate incorrect or insecure code. Sandboxed execution mitigates risk during evaluation. Deployment requires human oversight, static analysis, and audit mechanisms.


\bibliographystyle{IEEEtran}
\bibliography{ref}
\end{document}